# Lamb-dip spectroscopy of rotational levels with UTC-PD terahertz emitter

KOHEI EGUCHI,[1,6] TAKASHI ARIKAWA,[1,2,3] HIROSHI ITO,[4] AND KOICHIRO TANAKA[1,5,7]

[1] *Department of Physics, Graduate School of Science, Kyoto University, Sakyo-ku, Kyoto 606-8502, Japan.*
[2] *PRESTO, Japan Science and Technology Agency (JST), Saitama, Japan.*
[3] *Present address: Graduate School of Engineering, University of Hyogo, Himeji, Japan.*
[4] *Institute for Photon Science and Technology, Graduate School of Science, The University of Tokyo, Bunkyo-ku, Tokyo, Japan.*
[5] *Research Center for Advanced Photonics, RIKEN, 2-1 Hirosawa, Wako, Saitama, 351-0198, Japan.*
[6] *eguchi.kohei.f95@kyoto-u.ac.jp*
[7] *koichiro.tanaka@riken.jp*

**Abstract:** Pump-probe saturation spectroscopy in the sub-terahertz region was performed in the rotational transition $(J, K) = (16, 0) \leftarrow (15, 0)$ for gas-phase acetonitrile molecules in the counter-propagating configuration. We observed Lamb-dips at much lower excitation powers than previously reported. The linewidth in the zero-pressure limit was 10 kHz, which was estimated from the intensity and pressure dependence. This corresponds to the transit-time broadening.

## 1. Introduction

The rotational transition of unipolar molecules gives rise to molecular-specific spectra, and has been used in chemical analyses and searches for interstellar molecules [1,2]. Presently, molecular spectroscopic parameters (e.g., absolute transition frequencies) are catalogued in the JPL database [3] and HITRAN [4]. However, most of them are based on the linear absorption region, where the absorption linewidth is dominated by the Doppler inhomogeneous width resulting from thermal motion of the molecules [5,6]. The accuracy was constrained to $10^{-9}$. The rotational levels of molecules are not equally spaced, thus allowing them to be treated as a two-level system. This has led to the development of molecular clocks based on atomic clocks [7,8]. The background to this kind of research is the establishment of a frequency standard in the microwave to terahertz (THz) range using rotational transitions, which has not yet been achieved [9]. The necessity for such a standard has grown with the emergence of sixth-generation (6G) communications [10] and THz radar [11]. To develop a standard, it is necessary to lock the laser frequency to the rotational transition of the molecule. Frequency locking to rotational transitions has already been achieved in several types of terahertz oscillator, including quantum cascade lasers (QCL) [12,13], complementary metal-oxide-semiconductor (CMOS) circuits [14], and photomixers [15,16,17]. However, since linear absorption was used in these studies, the Doppler inhomogeneous width imposed constraints on the phase noise and absolute frequency accuracy [5,6]. Lamb-dip spectroscopy has emerged as a promising solution to these issues [6,18,19]. Using this spectroscopy, narrow dips can be obtained at the positions of the absolute transition frequencies in the Doppler-inhomogeneity-broadened spectrum. Therefore, by using this Lamb-dip to perform frequency locking, it is possible to determine the absolute frequency with high precision, and it is expected that phase noise can be reduced. However, only a few studies have been performed to determine the absolute frequency of pure rotational transitions by using Lamb-dip spectroscopy [20-22] and to lock the laser frequency [23]. This is due to the lack of narrow-linewidth terahertz emitters with tunable wavelengths and sufficiently sensitive detection methods that can be used for Lamb-dip spectroscopy. As a result, it is still unclear to what extent the absorption saturation behavior of rotational transitions can be described by a theoretical model based on the two-level system approximation. In particular, since there has been no systematic experimental study of absorption saturation, the discussion of saturation intensities obtained from experiments and theory is insufficient.

Here, further improvements in laser frequency locking can be expected by decreasing the homogeneous width. At the pressure of the gas used in Lamb-dip spectroscopy, the observed homogeneous width is mainly governed by collision broadening and saturation broadening [5], but these aspects have not been systematically examined yet. Furthermore, the reduction of the homogeneous width allows us to resolve hyperfine splitting such as nuclear spin splitting [24]. Therefore, it is very important to investigate the homogeneous width observed in Lamb-dip spectroscopy in detail.

Here, we performed Lamb-dip spectroscopy on the ($J$, $K$)=(16, 0)←(15, 0) rotational transition of acetonitrile molecules by using a uni-traveling carrier photodiode (UTC-PD) [25], which is a typical low-power THz emitter. The UTC-PD is an optoelectronic device that generates THz light by taking the difference between two frequencies of light, and its advantage is that it can be precisely controlled optically. We have generated frequency-stabilized THz light by using two lasers frequency-locked to an optical comb [26]. By systematically examining the dependence of the saturated absorption on sample pressure and irradiation intensity, we found that the saturation intensities obtained from the two-level system approximation are in good agreement with those obtained experimentally. We succeeded in observing Lamb-dip at a power of only 8 μW. By extrapolating the intensity dependence of the homogeneous width obtained by Lamb-dip spectroscopy, it was determined that the homogeneous width is proportional to the pressure in the zero-intensity limit and is limited to transit-time broadening in the zero-pressure limit.

**2.  Experiment**

## 2.1 Sample selection

In the case of gas-phase molecular saturation spectroscopy, an important parameter for molecular selection should be a large transition dipole moment. This allows for highly accurate absorption measurements with large oscillator strength and small saturation intensities.

Since the rotational levels are not equally spaced, we can use a two-level system approximation to estimate the saturation intensity $I_s$ [27-29] as follows:

$$I_s = 3c\varepsilon_0 \hbar^2 \tilde{\gamma}^2 / (2|\mu_{ij}|^2) , \qquad (1)$$

where $c$ is the speed of light, $\varepsilon_0$ is the electric permittivity in free space, $\hbar$ is the reduced Planck constant, $\tilde{\gamma}$ is the homogeneous width (rad·Hz) corresponding to the half-width at half-maximum (HWHM) of the Lorentzian profile, and $|\mu_{ij}|$ is the magnitude of the transition dipole moment. We assumed that the homogeneous width is determined by two dominant processes that contribute equally to the population decay and the polarization decay [30,31]. Under these conditions, the homogeneous width is expressed as $\tilde{\gamma} = 2\pi\gamma_p + 2\pi\gamma_{tt}$, where the former term is due to collision broadening (Hz) [5], $\gamma_p = p \times C_p$, and the latter is due to transit-time broadening (Hz) [5], $\gamma_{tt} = (u/(2\pi w))\sqrt{2\ln 2}$. Here, $p$ is the pressure, $C_p$ is collision broadening coefficient, $u$ is the mean thermal velocity, and $w$ is beam radius.

Table 1 lists $|\mu_{ij}|^2, C_p, \gamma_{tt}$ at $w$=4.11 mm, the oscillator strength $f_{ij}$, and the saturation intensity for three molecules that are frequently used in THz rotational spectroscopy in the sub-THz region. Here, we assumed a sample pressure of 0.1 Pa. We selected the acetonitrile molecule ($CH_3CN$) for its large oscillator strength and small saturation intensity.

**Table 1.** Comparison of square of transition dipole moment $|\mu_{ij}|^2$, collision broadening coefficient $C_p$, transit-time broadening $\gamma_{tt}$, oscillator strength $f_{ij}$, and saturation intensity $I_s$ at 0.1 Pa.

| Molecules | $|\mu_{ij}|^2$ (Debye²)[a] | $C_p$ (kHz/Pa)[b] | $\gamma_{tt}$ (kHz) | $f_{ij}$ (−)[c] | $I_s$ (mW/cm²) at 0.1 Pa |
|---|---|---|---|---|---|
| $CH_3CN$ | 7.94 | 539(3)* | 8.9 | 3.90×10⁻⁵ | 0.00782 |
| OCS | 0.262 | 49.1 | 7.4 | 1.15×10⁻⁶ | 0.0091 |
| $N_2O$ | 0.00135 | 30.5 | 8.6 | 6.94×10⁻⁸ | 0.158 |

[a]Square of transition dipole moment is calculated by $|\mu_{ij}|^2 = \mu^2((J+1)^2 - K^2)/((J+1)(2J+1))$. Magnitude of permanent dipole moment, $\mu^2$, is from the JPL database [3]. [b]Collision broadening coefficient, $C_p$, is from HITRAN 2020 [4]. *Collision broadening coefficient of $CH_3CN$ is determined by rotational-vibrational spectroscopy [32]. [c]The oscillator strength is calculated by $f_{ij} = (2m_e\omega_0)/(3e^2\hbar)(g_2/g_1)|\mu_{ij}|^2$ [33], where $m_e$ and $e$ are mass of a free electron and the elementary charge, respectively. $\omega_0$ is the transition frequency. $g_2$ and $g_1$ are degeneracies of the upper and lower level. All table values are based on the (J, K)=(16, 0)←(15, 0) transition for $CH_3CN$, the J=22←21 transition for OCS, and the J=12←11 transition for $N_2O$ [4].

## 2.2 Rotational levels of acetonitrile molecules

The acetonitrile molecule has a symmetric-top structure, as illustrated in Fig. 1(a). The energy of the rotational levels is given by [24]

$$E(J) = hBJ(J+1) + h(A-B)K^2 - hD_J J^2(J+1)^2 - hD_{JK} J(J+1)K^2 - hD_K K^4, \quad (2)$$

where $J$ is the total angular momentum of the two orthogonal rotational modes, and $K$ is the $z$-component of the total angular momentum. $J$ takes an integer value greater than or equal to 0; thus, $K$ takes an integer value between $-J$ and $+J$. Equation (2) degenerates for positive and negative $K$: thus, a only positive $K$ is treated below. $A$ and $B$ are determined from the inverse of momentum of inertia parallel and perpendicular to the symmetric axis, respectively. $D_J, D_{JK}$ and $D_K$ are determined from centrifugal distortion. $h$ is the Planck constant. Figure 1(b) shows the energy diagram of the $J$=15 and $J$=16 rotational levels. The parameters for the calculation are summarized in Table 2. As can be seen in Fig. 1(b), the magnitude of the splitting due to $K$ is one order of magnitude larger than that due to $J$. On the other hand, the rotational transition selection rules are $\Delta J = \pm 1$ and $\Delta K = 0$; therefore, the allowed absorption transition is $(J+1, K) \leftarrow (J, K)$ (Black arrows in Fig. 1(b)). From eq. (2) and the selection rules, the transition frequency is given by

$$v_0(J+1, K \leftarrow J, K) = 2B(J+1) - 2D_{JK}(J+1)K^2 - 4D_J(J+1)^3. \quad (3)$$

The difference in transition frequency due to $J$ is roughly determined as $2B$. The simulated absorption spectrum of the $(16, K) \leftarrow (15, K)$ transitions at 295 K is shown in Fig. 1(c). One can see that a small splitting by $K$ takes place that is due to the $-2D_{JK}(J+1)K^2$.

**Table 2.** Ground state spectroscopic parameters of acetonitrile [34].

| | |
|---|---|
| $A - B$ | 148.900074(65) GHz |
| $B$ | 9.198899134(11) GHz |
| $D_{JK}$ | 177.40796(28) kHz |
| $D_J$ | 3.807528(9) kHz |
| $D_K$ | 2.8251(15) MHz |

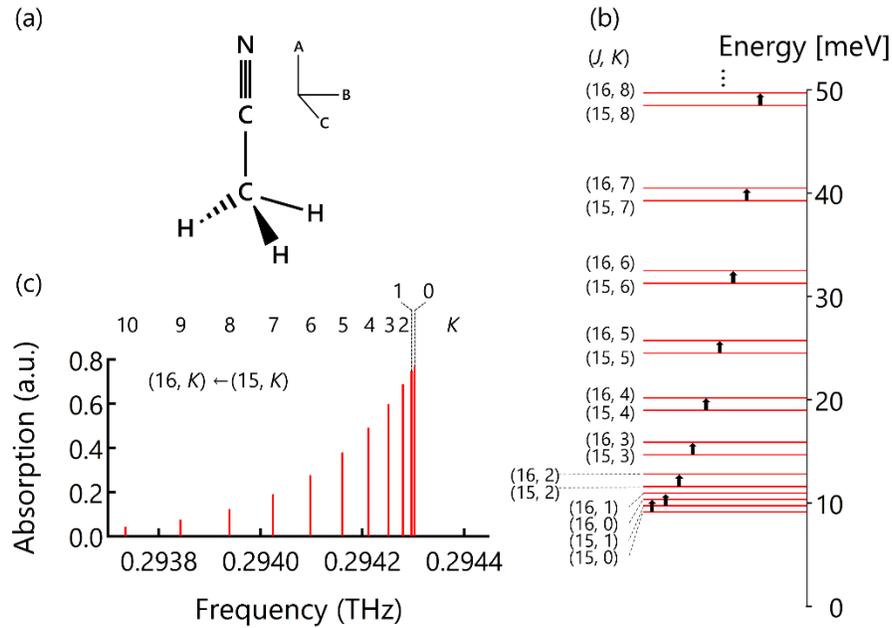

**Fig. 1.** (a) Molecular structure of acetonitrile. (b) The energy diagram of $J=15$ and $J=16$ rotational levels of acetonitrile molecule calculated with eq. (2). Black arrows show the allowed transitions. (c) Simulated absorption spectrum of $(J, K)=(16, K)\leftarrow(15, K)$ transition of acetonitrile molecules at 295 K.

## 2.3 Experimental setup

Figure 2(a) shows a schematic layout of the transmission spectroscopy. The THz emission from the uni-traveling carrier photodiode (UTC-PD, IOD-PMJ-13001, NTT Electronics, Co.) [25] was split by the Si beam splitter (Si-BS1). The reflected light (red arrow) was detected by the Fermi-level managed barrier diode (FMBD1) [35] after passing through the gas cell (10 cm long, 25 mm diameter), and the transmitted light (blue arrow) was detected by another Fermi-level managed barrier diode (FMBD2) as a reference signal. We used the FMBDs in power detection mode. The THz frequency was scanned using the method described below.

Figure 2(b) shows the schematic layout of the Lamb-dip spectroscopy. The counter-propagating pump-probe configuration was realized by simply rotating wire-grid polarizer 2 (WGP2) and WGP4 by 90° from the angles in the layout shown in Fig. 2(a). The reflected light at Si-BS1 (red arrow) pumped the molecules and the transmitted light (blue arrow) probed the excited molecules. The total THz power was controlled by adjusting the input lasers power of the UTC-PD, while the probe power was kept constant by rotating WGP3. The lights propagated colinearly. The beam radius at the center of the gas cell was $w=4.11$ mm. All WGPs were slightly tilted to prevent cavity formation. The gas cell windows were wedged. The sample was flowed very slowly, and the pressure was monitored by an attached pressure gauge. The pressure was corrected by comparing the ionization cross sections of acetonitrile and nitrogen molecules [36].

Figure 2(c) shows a block diagram of the frequency stabilization and control of the two lasers (DLpro and CTL1550, TOPTICA Photonics AG) working around 1550 nm for THz generation. Both lasers were highly frequency-stabilized by phase locking to the longitudinal modes of an optical-comb (OCLS-100DP-KY, NEOARK CORPORATION). Small parts of the laser outputs were coupled with the optical-comb and used to generate error signals for the feedback controllers (OCLS-STB-KY, NEOARK CORPORATION). The laser frequencies were stabilized so that the beats of the individual lasers and neighboring comb modes were maintained at the externally supplied lock-offset frequency $f_{\text{Lock-offset}}$, typically set to 10 MHz (Fig. 2(d)). Therefore, the laser frequency is expressed as $f_{\text{laser}} = f_{\text{CEO}} + n\, f_{\text{rep}} + f_{\text{Lock-offset}}$. Here, $n$ is the index of the comb mode, and $f_{\text{CEO}}$ and $f_{\text{rep}}$ are the carrier envelope offset and repetition frequency of the optical-comb, respectively. $f_{\text{CEO}}$ and $f_{\text{rep}}$ were set at 10 MHz and 100 MHz by referencing the global positioning satellite (GPS) signal. Two lasers were delivered to the UTC-PD through polarization maintaining (PM) fiber, and excited the UTC-PD. The resulting THz frequency is expressed as

$$f_{\text{THz}} = |f_{\text{Laser1}} - f_{\text{Laser2}}| = |(n_1 - n_2)f_{\text{rep}} + (f_{\text{Lock-offset1}} - f_{\text{Lock-offset2}})|. \quad (4)$$

$n_1$ and $n_2$ were set at 1936330 and 1933387, respectively, to be closest to the target absorption line. The THz frequency was swept by sweeping both lock-offset frequencies. The THz linewidth was less than 120 mHz [26]. The laser was modulated at 10 kHz prior to exciting the UTC-PD. The signals from two FMBDs were acquired by the lock-in amplifier at the same time.

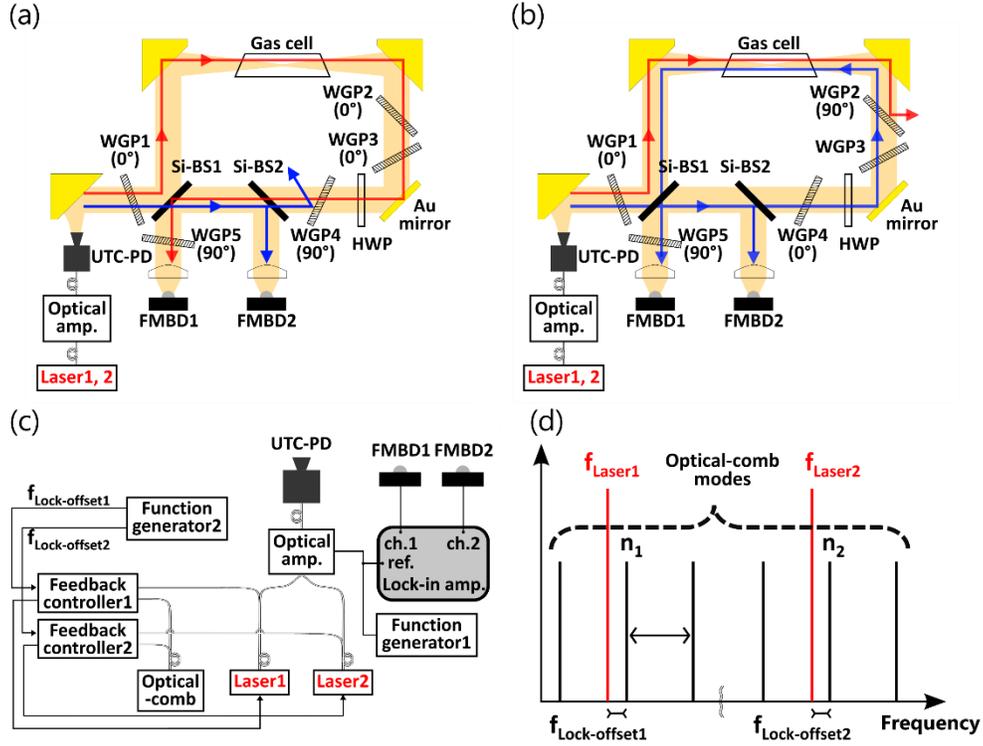

**Fig. 2.** (a) Schematic layout of transmission spectroscopy. UTC-PD: Uni-traveling carrier photodiode, FMBD: Fermi-level managed barrier diode, WGP: Wire-grid polarizer, HWP: Half wave plate, Si-BS: Si beam splitter. (b) Schematic layout of the Lamb-dip spectroscopy. WGP2 and WGP4 are rotated 90° from the transmission layout. (c) Block diagram of laser frequency control and data acquisition circuit. The black lines are electrical lines, and the gray lines are optical fibers. (d) Schematics diagrams of laser frequency stabilization and control method. Black lines and red lines are comb modes and lasers frequency, respectively.

## 3. Results

The rotational absorption spectrum is derived from the optical Bloch equations under the two-level system approximation and Doppler inhomogeneous width and is expressed by the Voigt function [5,27-29],

$$\alpha_0(\omega) = -\int_{-\infty}^{\infty} dv_z \frac{\exp(-v_z^2/v_p^2)}{v_p\sqrt{\pi}} \times \frac{2|\mu_{ij}|^2}{3\varepsilon_0 c\hbar} \frac{\omega\tilde{\gamma}}{(\omega_0 - \omega - kv_z)^2 + \tilde{\gamma}^2} N\eta , \qquad (5)$$

where $\omega$ is the angular frequency, $\omega_0$ is the transition frequency, $v_p$ is the most probable velocity of thermally distributed molecules, $v_z$ is the molecular velocity component parallel with the light propagation, $N$ is the number density of molecules, and $\eta$ reflects the initial thermal population based on the Maxwell-Boltzmann distribution [24,29]. The saturation effect is ignored. The absorption linewidth corresponding to the Voigt half-width is determined from a Lorentzian profile with a homogeneous width $\tilde{\gamma}$ and Doppler inhomogeneous width $\delta\omega_D =$

$2\sqrt{\ln 2}(2\pi\nu_0)v_p/c$, which is induced by molecular thermal motion. It is clear that at high pressure, where the collision broadening is sufficiently larger than the Doppler width, the homogeneous width dominates the absorption linewidth. At low pressures, where the collision broadening is sufficiently smaller, the Doppler width dominates the absorption linewidth.

### 3.1 Transmission spectroscopy at relatively high pressure

First, wide-range transmission spectroscopy was performed on rotational absorption of acetonitrile molecules to obtain an overview of the spectrum under slightly reduced pressure conditions with the layout shown in Fig. 2(a). Here, we used a wavelength meter (Angstrom WS7/30 IR, Highfinesse GmbH) to stabilize and control the laser frequencies instead of the optical-comb, since the frequency control using the optical-comb can sweep only 7 MHz, which is much smaller than rotational level spacing ($2B$ ~18.4 GHz). Figure 3(a) shows the absorption spectra measured at 5 kPa. A comparison with the JPL database [3] identified five peaks corresponding to the transitions $J = 14 \leftarrow 13, J = 15 \leftarrow 14, J = 16 \leftarrow 15, J = 17 \leftarrow 16$, and $J = 18 \leftarrow 17$. The blue lines in the figure show the absorption frequencies listed in the database. The peaks are mostly aligned with a $2B$ spacing corresponding to the first term of the right-hand side of eq. (3). The small oscillating component (period=700 MHz) is the remaining etalon effect. The $K$ splitting structure is not visible in the broad absorption lines at such high pressure. Figure 3(b) shows an enlargement of the spectrum around the $J = 16 \leftarrow 15$ transition measured at 10 Pa. Here, the $K$ splitting structure is resolved because the collision broadening is reduced to 5.4 MHz. The components with $K$=2 to 10 were well decomposed and matched the database [3], while those with $K$=0 and $K$=1 could not be resolved. Figure 3(c) shows an enlarged spectrum around $K = 0$ and $K = 1$ transitions at 3 Pa measured by fine frequency control with the optical-comb. Here, the $K = 0$ and $K = 1$ absorptions are well resolved and were reproduced by the Voigt function. Thus, for acetonitrile at a sufficiently low-pressure below 3 Pa, collision broadening is reduced and is comparable with the Doppler width (564.96 kHz).

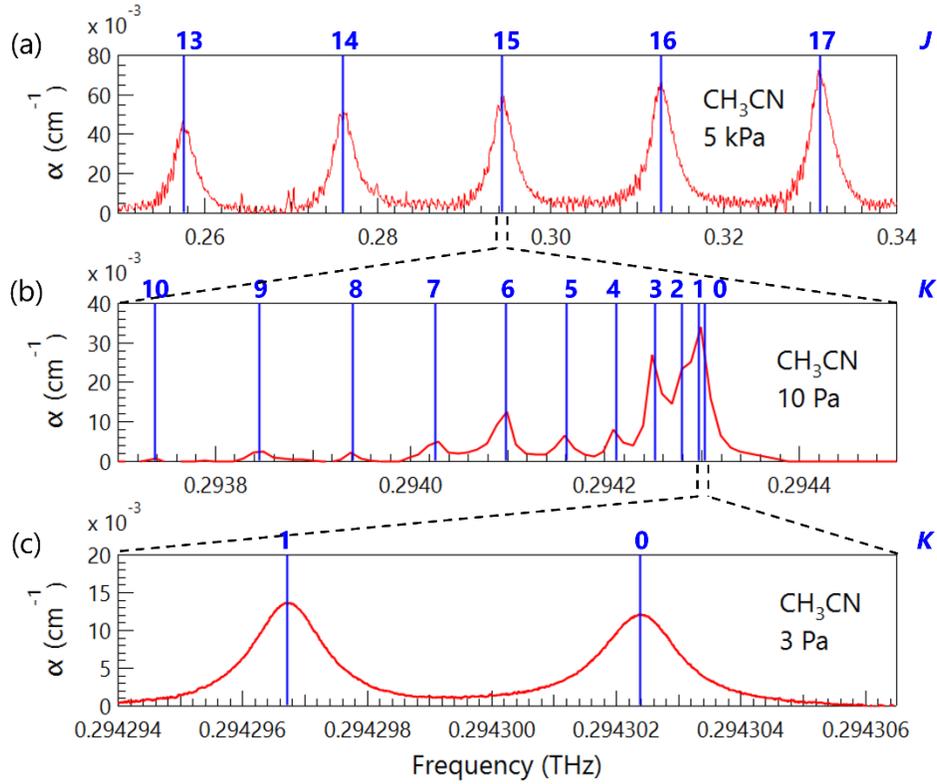

**Fig. 3.** Absorption spectra of acetonitrile molecules measured at relatively high-pressure. (a) Spectra measured at 5 kPa and (b) at 10 Pa measured with frequency control by wavelength meter and (c) at 3 Pa with frequency control using an optical-comb. Red line: experiment, blue line: JPL database.

### 3.2 Pressure dependence of the absorption at sufficiently low-pressure

At pressures where the collision broadening is sufficiently smaller than the Doppler inhomogeneous width, the homogeneous width is buried within the Doppler width. Even in this case, however, we can obtain the homogeneous width from a fitting by the Voigt function (eq. (5)).

Figure 4(a) shows absorption spectra of acetonitrile molecules for the $(J,K) = (16, 0) \leftarrow (15, 0)$ rotational transition at sample pressures of 0.25 Pa (red dots), 0.15 Pa (green dots) and, 0.1 Pa (blue dots). The horizontal axis is the frequency deviation from the absorption center $v_0$ which was determined by Lamb-dip spectroscopy as described in section 3.4. We performed a global fitting where several spectra measured at various pressures were fitted simultaneously with Voigt functions sharing the Gaussian width and absorption center frequency. The black curves are the best-fitted results. Figure 4(b) shows the pressure dependence of the homogeneous width, $\tilde{\gamma}_{exp}/2\pi$ (red circles) and absorption linewidth, $\gamma_{abs,exp}$ (blue circles), both obtained by fitting. The red curve is the theoretically calculated homogeneous width given by $\tilde{\gamma}/2\pi = \gamma_p + \gamma_{tt} = (539.4 \times \text{pressure} + 8.9)$ kHz, as described in section 2.1. The blue curve is the calculated absorption linewidth given by $\gamma_{abs} = \tilde{\gamma}/4\pi + \sqrt{(\tilde{\gamma}/4\pi)^2 + (\delta\omega_D/4\pi)^2}$. The Doppler half-width of the $(J, K) = (16, 0) \leftarrow (15, 0)$ transition at room temperature is $\delta\omega_D/4\pi = \sqrt{\ln 2}\, v_0 v_p/c = 282.48$ kHz. The absorption linewidth decreases with decreasing sample pressure, and becomes asymptotic to the Doppler half-width. Even with the collision broadening coefficient obtained from rotational-vibrational transitions [32] (Table 1), theoretical calculation reproduces the experimental results well. Figure 4(c) shows the pressure dependence of the experimentally obtained maximum value of the absorption peak (red circles). The red curve plots the theoretically calculated absorption maxima at $\omega = \omega_0$ in eq. (5). Although the error is large at low pressure, the experimental and theoretical values are good agreement.

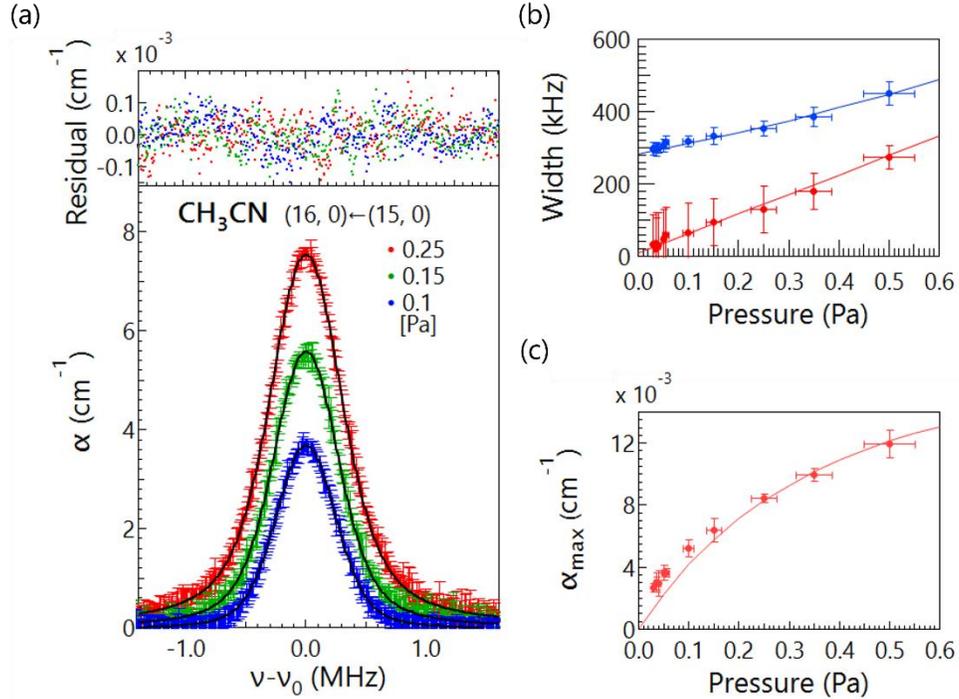

**Fig. 4.** (a) Pressure-dependent absorption spectra of acetonitrile molecules at low pressure. Bottom panel: Measured absorption spectra with 0.001 mW/cm²; red dots: 0.25 Pa, green dots: 0.15 Pa, blue dots: 0.1 Pa, black curve: Voigt fitting, Top panel: residuals between measured absorption spectra and fitting. The origin of the horizontal axis is the absorption center ($v_0 = 294302.389$ MHz) determined by Lamb-dip spectroscopy (section 3.4). (b) Pressure-dependent absorption linewidth (HWHM) and homogeneous width of acetonitrile. Blue circles: measured absorption linewidth, red circles: homogeneous width extracted by Voigt fitting, blue curve: calculated absorption linewidth, red curve: calculated homogeneous width. (c) Pressure-dependent absorption maximum. Red circle: measured absorption maximum, red curve: theoretically calculated absorption maximum.

### 3.3 Power dependence of the absorption at sufficiently low-pressures

So far, we have ignored the saturation effect. However, at sufficiently low-pressure, the homogeneous width tends to saturate significantly. Therefore, we examined the saturation effect carefully to determine the collision broadening coefficient of the pure rotational transition and saturation intensity.

In the two-level system approximation, saturation effects appear at sufficiently strong light intensities, and the homogeneous width broadens as follows [5,29]

$$\tilde{\gamma}_B = \tilde{\gamma}\sqrt{1+S} = \tilde{\gamma}\sqrt{1+I/I_s} \ , \tag{6}$$

where $I$ is the irradiation intensity, and $I_s$ the saturation intensity. The homogeneous width $\tilde{\gamma}$ in eq. (5) is replaced by this saturated homogeneous width $\tilde{\gamma}_B$. Experimentally, the saturation intensity and the unsaturated homogeneous width can be obtained by fitting the irradiation intensity dependence and pressure dependence of the saturated homogeneous width by using eq. (6).

Figure 5(a) shows the absorption spectra of acetonitrile molecules for the $(J, K) = (16, 0) \leftarrow (15, 0)$ rotational transition at 0.1 Pa with an excitation intensity of 0.001 mW/cm$^2$ (red dots), 0.02 mW/cm$^2$ (green dots,) and 0.06 mW/cm$^2$ (blue dots). We performed a global fitting where several spectra measured at various intensities were fitted simultaneously with Voigt functions sharing Gaussian width and absorption center frequency. The black curves are the best-fitted results.

Figure 5(b) shows the intensity dependence of the saturated homogeneous width, $\tilde{\gamma}_{B,exp}$, obtained by the global fitting. The green circles are results for 0.5 Pa, orange circles for 0.35 Pa, pink circles for 0.2 Pa, purple circles for 0.15 Pa, blue circles for 0.1 Pa, red circles for 0.05 Pa. One can see that saturation takes place largely in the low-pressure region. This should come from the small saturation intensity: the saturation intensity is described by $I_s = p^2/C_s$, $C_s = 2|\mu_{ij}|^2/(3c\varepsilon_0\hbar^2 C_p^2)$ from eq. (1), where the transit-time broadening is negligible. We made a global fitting of the saturated homogeneous width to $\tilde{\gamma}_B = \tilde{\gamma}\sqrt{1+I \times C_s/p^2}$ with fitting parameters $\tilde{\gamma}$ and $C_s$, with $C_s$ being shared.

The solid curves in Fig. 5(b) are the best-fitted results. The saturation intensities obtained by fitting and by the theoretical calculation for several pressures are summarized in Table 3. The theory reproduced the values within the error. The intercepts give the unsaturated homogeneous width at the zero-intensity limit. The collision broadening coefficient is estimated to be 529(10) kHz/Pa by a linear fitting to the pressure dependence of the unsaturated homogeneous width. This value is in close agreement with the value obtained by rotational-vibrational spectroscopy (Table 1) [32].

Figure 5(c) shows the irradiation intensity dependence of the peak value of the absorption coefficient. As expected, the peak value increased with increasing pressure. This mainly comes from the increase in molecular density. The decrease in the peak value is significant at low pressure because of the larger saturation broadening. The solid curves in Fig. 5(c) are not fitted results, but the absorption maxima calculated using eq. (5) with the saturated homogeneous widths obtained above. They reproduce the experimental results and guarantee the validity of our theoretical treatment.

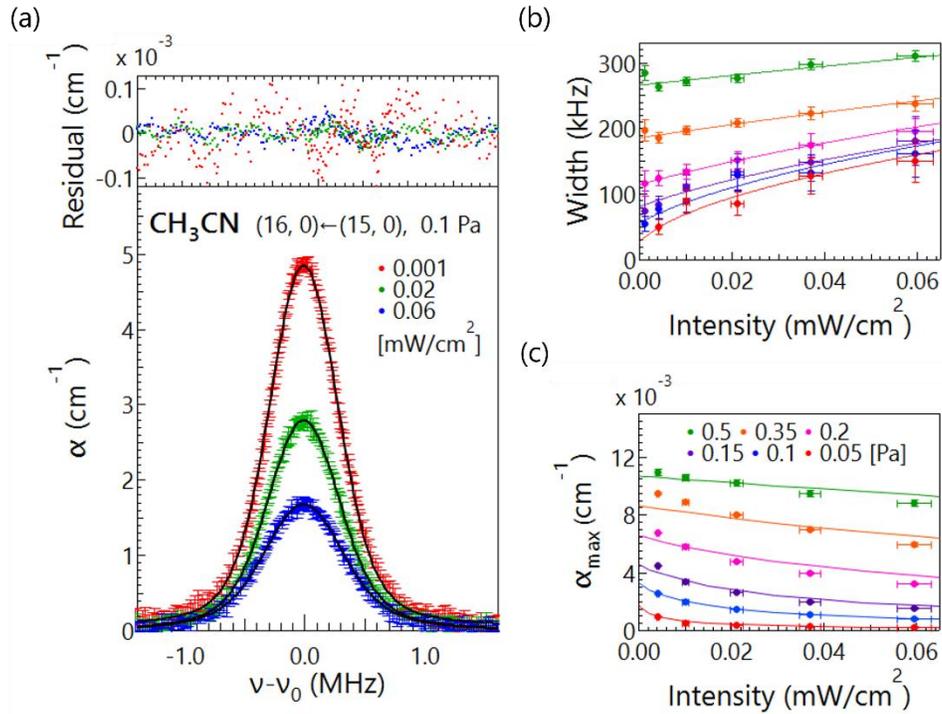

**Fig. 5.** (a) Power dependence of absorption spectra of acetonitrile molecules. Bottom panel: Measured absorption spectra at 0.1 Pa; red dots: 0.001 mW/cm², green dots: 0.02 mW/cm², blue dots: 0.06 mW/cm², black curve: Voigt fitting with saturation width, Top panel: residuals between measured absorption spectra and fitting. The origin of the horizontal axis is the absorption center (294302.389 MHz) determined by Lamb-dip spectroscopy (section 3.4). (b) Power dependence of homogeneous width of acetonitrile: green circles measured at 0.5 Pa, orange circles 0.35 Pa, pink circles 0.2 Pa, purple circles 0.15 Pa, blue circles 0.1 Pa, red circles 0.05 Pa. The solid curves are fitting with eq. (6). (c) Pressure dependence of maximum absorption coefficient: green circles measured at 0.5 Pa, orange circles 0.35 Pa, pink circles 0.2 Pa, purple circles 0.15 Pa, blue circles 0.1 Pa, red circles 0.05 Pa. The solid curves are absorption maxima calculated using the estimated saturation intensity.

**Table 3.** Theoretical and experimental saturation intensities $I_s$ (mW/cm$^2$) of acetonitrile molecule for the $(J, K)$=(16, 0)←(15, 0) rotational transition.

| Pressure (Pa) | Theory (mW/cm$^2$) | Experiment (mW/cm$^2$) |
|---|---|---|
| 0.5 | 0.154 | 0.17(3) |
| 0.35 | 0.0773 | 0.086(14) |
| 0.2 | 0.0270 | 0.028(4) |
| 0.15 | 0.0160 | 0.016(3) |
| 0.1 | 0.00782 | 0.007(1) |
| 0.05 | 0.00255 | 0.002(1) |

### 3.4 Lamb-dip spectroscopy

We performed Lamb-dip spectroscopy in the pressure range below 0.1 Pa, where the Doppler inhomogeneous width is dominant. The layout shown in Fig. 2(b).

The absorption spectrum for the Lamb-dip is described by eq. (7) below [18,37], where $\alpha_0(\omega)$ is the Doppler broadened unsaturated absorption given by eq. (5). The term in square brackets represents a Lorentz shape dip structure with a width of $\tilde{\Gamma}$ (eq. (8)) and peak at $\omega_0$. It should be noticed that the dip width, $\tilde{\Gamma}$ is determined only by the homogeneous width $\tilde{\gamma}$ and is affected by the saturation.

$$\alpha_{\text{Lamb}}(\omega) = \alpha_0(\omega)\left[1 - \left(1 - \frac{1}{\sqrt{1+S}}\right)\frac{\tilde{\Gamma}^2}{\tilde{\Gamma}^2 + (\omega - \omega_0)^2}\right] \quad (7)$$

$$\tilde{\Gamma} = \tilde{\gamma}(1 + \sqrt{1+S})/2 = \tilde{\gamma}\left(1 + \sqrt{1 + I/I_s}\right)/2 \quad (8)$$

Figure 6(a) shows Lamb-dip spectra of acetonitrile molecules for the $(J, K) = (16, 0) \leftarrow (15, 0)$ rotational transition measured at 0.1 Pa at three pump intensities: no pump (red dots), 0.02 mW/cm$^2$ (green dots) and, 0.06 mW/cm$^2$ (blue dots). Here, we performed a global fitting where several spectra measured at various pump intensities were fitted simultaneously using eq. (7) with a shared Gaussian width and absorption center frequency. The black curves are the best-fitted results. Since the dip width is the sum of the unsaturated homogeneous width and saturated homogeneous width (eq. (8)), the unsaturated homogeneous width should be estimated by fitting the pump intensity dependence. Figure 6(b) shows the dependence of the pump intensity on the dip width $\tilde{\Gamma}$ and results of a fitting by eq. (8). Green circles denote results for 0.15 Pa, orange circles 0.1 Pa, purple circles 0.075 Pa, blue circles 0.05 Pa, and red circles 0.03 Pa. The intercept gives the unsaturated homogeneous width in the zero-pump limit. Figure 6(c) shows the pressure dependence of the unsaturated homogeneous width, $\tilde{\gamma}_{exp}$. The red and black circles are the unsaturated homogeneous widths obtained from Fig. 5(b) and Fig. 6(b), respectively. The black line is obtained by linear fitting of the homogeneous width obtained by Lamb-dip spectroscopy. The slope and intercept correspond to the collision broadening coefficient $C_p$, 545(7) kHz/Pa, and transit-time broadening $\gamma_{tt}$, 8(1) kHz. The previously reported (Table 1) and theoretically calculated values of $C_p$ and $\gamma_{tt}$ are within the error of these measurements.

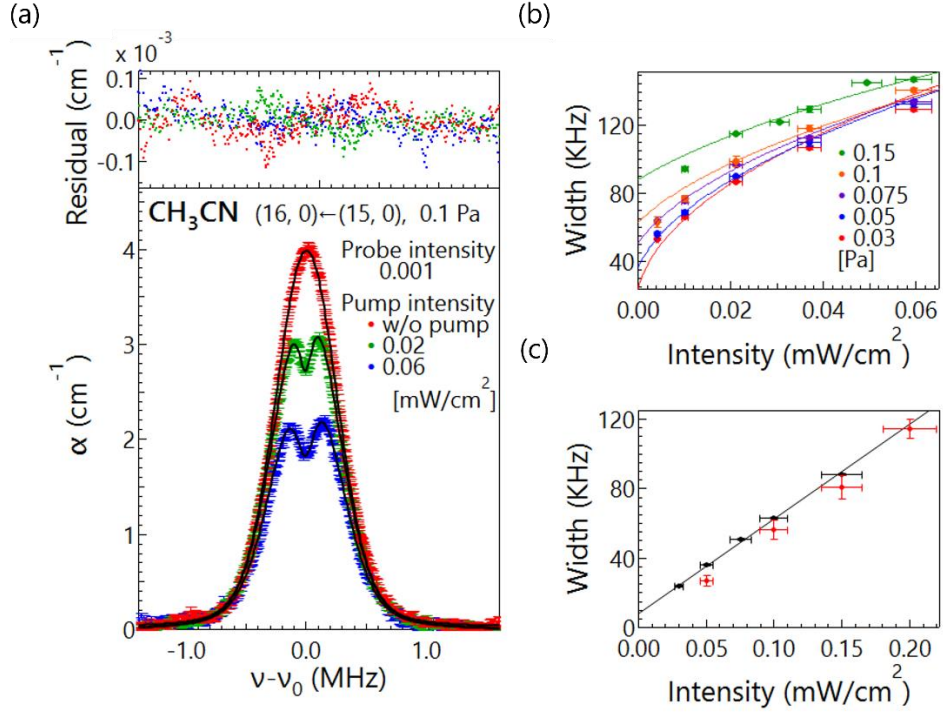

**Fig. 6.** (a) Pump-intensity dependence of Lamb-dip spectra of acetonitrile molecules at 0.1 Pa. Bottom panel: measured Lamb-dip spectra, red dots: w/o pump, green dots: 0.02 mW/cm², blue dots: 0.06 mW/cm², black curve: Lamb-dip fitting (eq. (7)), Top panel: residuals between measured absorption spectra and fitting. The origin of the horizontal axis the absorption center (294302.389 MHz) determined by Lamb-dip spectroscopy. (b) Pump-intensity dependence of dip width (HWHM) of acetonitrile. Green circles are results for 0.15 Pa, orange circles 0.1 Pa, purple circles 0.075 Pa, blue circles 0.05 Pa, and red circles 0.03 Pa, Solid curve is a fitting with eq. (8). (c) Pressure dependence of homogeneous width. Red circles are homogeneous widths obtained from Fig. 5(b), and black circles are homogeneous widths obtained from Fig. 6(b). The black line is a linear fitting of the homogeneous width obtained by Lamb-dip spectroscopy.

**3.5 Determination of the absolute transition frequency**

Finally, we discuss the absolute frequency of the rotational transition obtained by Lamb-dip spectroscopy stabilized with the optical-comb. Using the rotational level parameters reported previously [34] (Table 2), the frequency of the $(J, K) = (16, 0) \leftarrow (15, 0)$ transition is estimated as 294.30238975(38) GHz. In the current experiment, the transition frequency obtained by the transmission spectroscopy limited by the Doppler width is 294.302388(5) GHz (Fig. 4(a)). This value is an order of magnitude worse than the precision of the values estimated above. On the other hand, Lamb-dip spectroscopy gives 294.3023895(3) GHz (Fig. 6(a)). This value has the same accuracy as the previous report. This clearly demonstrates the usefulness of Lamb-dip spectroscopy.

4. **Summary and conclusion**

Transmission spectroscopy and Lamb-dip spectroscopy with a frequency-stabilized THz emitter were conducted on the rotational levels of the acetonitrile molecule in the THz region. We found that Doppler-free spectroscopy is possible even with a THz emitter with a power below 1 mW. The saturation intensity and homogeneous width were determined through a comprehensive measurement of the pressure and power dependence. Our findings indicate that these parameters can be successfully reproduced by a theoretical model based on a two-level system approximation. In a sufficiently low-pressure region, the homogeneous width is proportional to the pressure and reaches the transit-time broadening in the zero-pressure limit. The estimated collision broadening coefficient matches the previously reported value determined from the rotational-vibrational transition. The absolute frequency of the $(J, K) = (16, 0) \leftarrow (15, 0)$ transition as determined by the Lamb-dip spectroscopy is 294.3023895(3) GHz. This value is of the same accuracy as in the previous report, achieving a precision of $10^{-9}$.

Our report is expected to accelerate Lamb-dip spectroscopy of molecular rotational transitions, improve the accuracy of the databases, and enhance the performance of molecular clocks. For molecules with small saturation intensities, such as acetonitrile and OCS molecules, it is expected that a compact THz frequency standard can be realized using a resonant tunneling diode (RTD) [38]. Since the rotational levels of molecules can be treated as a two-level system and the dephasing time is long, they have been studied as potential qubits [39]. Moreover, the precise determination of the spectroscopic parameters and the resolution of the hyperfine structure of molecules would also make a great contribution.

**Acknowledgements**

This work was supported by JST ACCEL (Grant No. JPMJMI17F2), JSPS KAKENHI (Grant No. JP21H05017) and MEXT Q-LEAP (Grant No. JPMXS0118067634).